\documentclass{aastex63}

\def\red<#1>{\textcolor{red}{#1}}
\def\blue<#1>{\textcolor{blue}{#1}}

\received{August 9, 2019}
\revised{September 4, 2019}
\accepted{September 10, 2019}
\submitjournal{PASP}

\shorttitle{photometric precision of a mid-infrared detector}
\shortauthors{Matsuo et al.}

\begin{document}

\title{Photometric precision of a Si:As impurity band conduction mid-infrared detector and application to transit spectroscopy}

\correspondingauthor{Taro Matsuo}
\email{matsuo@iral.ess.sci.osaka-u.ac.jp}

\author{Taro Matsuo}
\affiliation{Department of Earth and Space Science, Osaka University, 1-1 Machikaneyama, Toyonaka, Osaka 560-0043, Japan}
\affiliation{NASA Ames Research Center, Moffett Field, CA 94035, USA}

\author{Thomas P. Greene}
\affiliation{NASA Ames Research Center, Moffett Field, CA 94035, USA}

\author{Roy R. Johnson}
\affiliation{NASA Ames Research Center, Moffett Field, CA 94035, USA}

\author{Robert E. Mcmurray}
\affiliation{NASA Ames Research Center, Moffett Field, CA 94035, USA}

\author{Thomas L. Roellig}
\affiliation{NASA Ames Research Center, Moffett Field, CA 94035, USA}

\author{Kimberly Ennico}
\affiliation{NASA Ames Research Center, Moffett Field, CA 94035, USA}

\begin{abstract}
Transit spectroscopy is the most promising path toward characterizing nearby terrestrial planets at mid-infrared (3 - 30 $\micron$) wavelengths in the next 20 years. The Spitzer Space telescope has achieved moderately good mid-infrared photometric precision in observations of transiting planets, but the intrinsic photometric stability of mid-IR detectors themselves has not been reported in the scientific or technical literature. Here, we evaluated the photometric precision of a James Webb Space Telescope (JWST) Mid-Infrared Instrument (MIRI) prototype mid-infrared Si:As impurity band conduction (IBC) detector, using time-series data taken under flood illumination. These measurements of photometric precision were conducted over periods of $\sim$ 10 hours, representative of the time required to observe an exoplanet transit. After selecting multiple sub-regions with a size of 10 x 10 pixels and compensating for a gain change caused by our warm detector control electronics for the selected sub-regions, we found that the photometric precision was limited to 26.3 $ppm$ at high co-added signal levels due to a gain variation caused by our warm detector control electronics. The photometric precision was improved up to 12.8 $ppm$ after correcting for the gain drift. We also translated the photometric precision to the expected spectro-photometric precision (i.e., relative photometric precision between wavelengths), assuming that an optimized densified pupil spectrograph is used in transit observations. We found that the spectro-photometric precision of an optimized densified pupil spectrograph when used in transit observations is expected to be improved by the square root of the number of pixels per a spectral resolution element. At the high co-added signal levels, the total noise could be reduced down to 7 $ppm$, which was larger by a factor of 1.3 than the ideal performance that was limited by the Poisson noise and readout noise. The systematic noise hidden behind the simulated transit spectroscopy was 1.7 $ppm$. 

\end{abstract}


\section{Introduction} \label{sec:intro}

Finding and characterizing the atmospheres of habitable planets orbiting nearby stars is one of the most important scientific goals for observatories that will be developed in the 2020s and 2030s. Near- and mid-infrared (IR) spectral features of ${\rm H_{2}O}$, ${\rm CH_{4}}$, ${\rm O_{3}}$, ${\rm CO_{2}}$, and other molecular species in the atmospheres of Earth-like planets are expected to have amplitudes of only 10 – 40 parts-per-million ($ppm$) in transmission or emission spectra when transiting mid-or-late M-dwarf stars \citep[e.g.,][]{2016MNRAS.458.2657B, 2016MNRAS.461L..92B, 2016ApJ...819L..13S}. Future observatories will have to produce observations with relative noise (measurement precision or stability) well under this value in order to detect these features. On the other hand, the residuals of the long-term phase curves taken by the Infrared Array Camera (IRAC) on the Spitzer Space Telescope (hereafter Spitzer) that applied Si:As impurity band conduction (IBC) detectors for 5.8 and 8 $\micron$ had standard deviation of 30 $ppm$ at high co-added signal levels \citep[e.g.,][]{2009ApJ...703..769K, 2012ApJ...754...22K}. The standard deviations of the residuals were not decreased by the square root of the number of the co-added integrations after some time period. \cite{2010ApJ...721.1861A} also acquired a similar result on the standard deviation of the residuals, after compensating the photometric variation caused due to the pointing jitter of Spitzer and removing a steep ramp of a few $\%$ at the beginning of the transit measurements (hereafter detector ramp). Note that the detector ramp could be also mitigated by exposing the detector to bright diffused light before beginning transit measurements, which is referred as to "pre-flash technique" \citep{2009ApJ...703..769K}. On the other hand, although the IRAC channels 1 and 2 (3.6 and 4.5 $\micron$) use In:Sb detectors, a different type of detector from that considered in this paper, the uncertainties of the transit depth measurements in the two channels taken during the post-cryogenic period of Spitzer were 60 $-$ 100 $ppm$ for XO-3b \citep{2016AJ....152...44I, 2016ApJ...820...86M}. In addition, \cite{2019arXiv190402294K} found that the temporal variations of the transit depth at 3.6 and 4.5 $\micron$ over four years were 20 $-$ 130 $ppm$ for HD 209458b and 90 $-$ 100 $ppm$ for HD 189733b. Although what limits the photometric precision of the Spitzer IRAC still remains unknown, one of the limitations is thought to be instrumental systematic noise.

A densified-pupil spectrograph forms spectra of pupil segments onto the detector and its signals are relatively stable to varying optical aberrations and small pointing errors \citep{2016ApJ...823..139M}. Furthermore, thanks to the spectra being spread over the entire detector, the local photometric variation generated by bad pixels, cosmic rays, and a detector control electronics is smoothed out through averaging a number of the pixels exposed by the object light  \citep{2018AJ....156..288G}. As a result, the relative photometric error between wavelengths is thought to be reduced for the densified pupil design. The number of pixels allocated for each resolution element is more than 1000 for the densified pupil spectrograph designed for the mid-infrared imager, spectrometer, coronagraph (MISC) of the Origins Space Telescope (OST) \citep{2018SPIE10698E..44M}. Thus, compared to the transit spectroscopy performed by the Spitzer and JWST \citep[e.g.,][]{2014PASP..126.1134B, 2016ApJ...817...17G}, OST/MISC is expected to be much less affected by instrument systematic noise.

On the other hand, the densified pupil design may not help to reduce the systematic noise  caused by the mid-infrared detector because the noise is caused by electrical variation in the detector system. In other words, the densified pupil design was originally proposed for improving the photometric variation caused by the optical aberration except for the image motion loss on the field stop \citep{2017AJ....154...97I}. No laboratory time-series photometric stability tests have been performed on the Si:As IBC detectors to date. We do have on-orbit time-series stability information from Spitzer \citep[e.g.,][]{2009ApJ...703..769K, 2012ApJ...754...22K}, but those were end-to-end system measurements and did not reveal the performance of the detector alone. We need to understand the time-series stability of the detector itself to determine how well we will be able to measure the mid-IR transmission, emission, and phase curve spectra of exoplanets with future space telescopes.

Based on this background, we measured the photometric stability of time-series data of a JWST MIRI prototype Si:As IBC detector that had been developed at NASA Ames Research Center \citep{2003SPIE.4850..890E}. In this study, we measured the photometric precision of the detector and used it to estimate the expected spectro-photometric precision of an optimized densified pupil spectrograph when used in transit observations. These multi-hour measurements have been conducted under ideal conditions; a cryogenic blackbody source was precisely controlled to within 1 $m$K and the detector was uniformly illuminated by a blackbody source. Therefore, the time-series data acquired in the measurements is thought to show the stability of the mid-infrared detector itself. This paper is organized as follows. In Section 2, we describe the acquisition and analysis of the Si:As IBC detector time-series data. In Section 3, we evaluate the achieved photometric precision and investigate its limitations. In Section 4, we translate the measured precision to the expected spectro-photometric precision of an ideal densified pupil spectrograph when used in transit spectroscopy. 

\section{Experiment} \label{sec:experiment}
In this section, we describe how we constructed time-series data from a number of uniformly illuminated images formed on the mid-infrared detector. After introducing the overview of the detector test in Section \ref{subsec:overview} and explaining the preparation of the data analysis in Section \ref{subsec:preparation}, we show the process for constructing the time-series data from images taken under flood illumination in Section \ref{subsec:gain}. The procedure includes selection of regions used for analyzing the time-series data and compensating the gain change observed in the time-series data. 

\subsection{Overview \label{subsec:overview}}
\begin{figure}[ht!]
\plotone{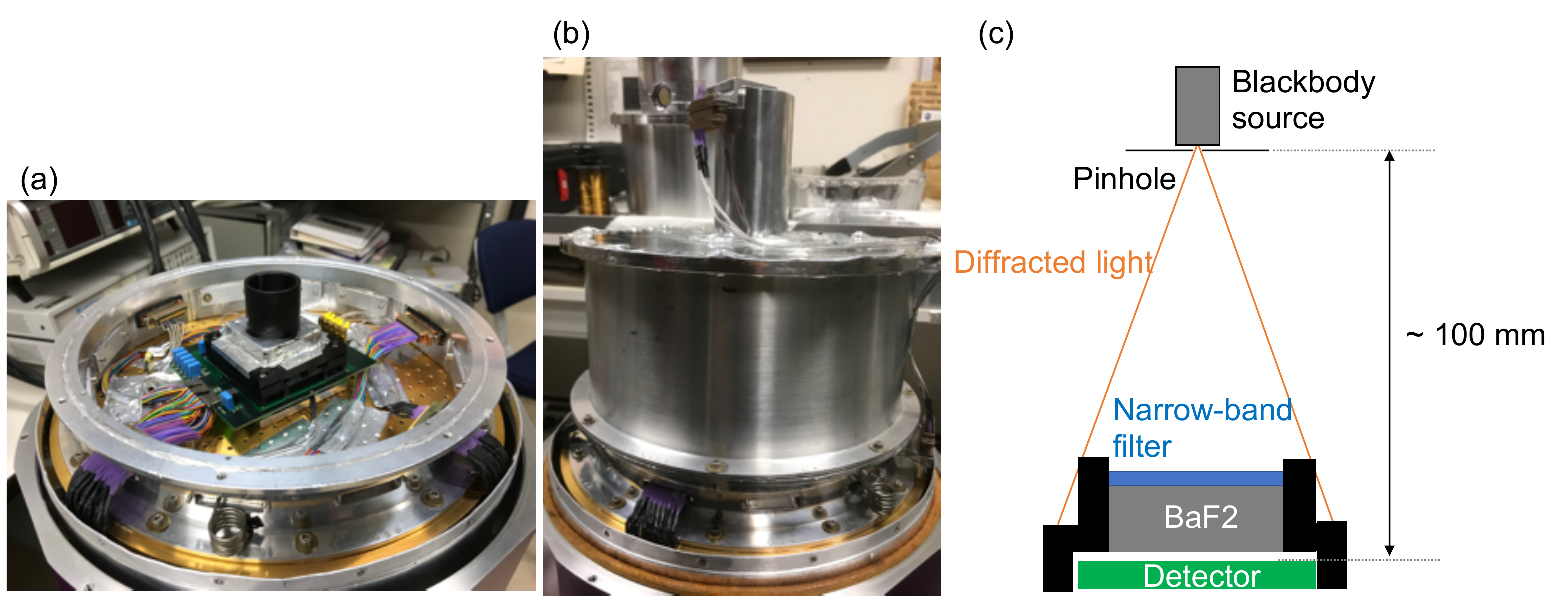}
\caption{(a) Picture of the detector system mounted on the cryogenic cold surface. (b) Picture of the detector test system covered by the aluminum Liquid Helium (LHe) shield. The silver cylinder attached to the upper rid of the LHe shield is the cryogenic blackbody source. (c) The schematic view of the detector optical test system.  \label{fig:configuration}}
\end{figure}

We acquired a number of datasets for measurement of the basic parameters and the stability for the bare readout integrated circuit (hereafter ROIC) and a hybridized Si:As IBC detector with SB226 ROIC architecture \citep{2003SPIE.4850..890E} from Dec. 7 through Dec. 20, 2018, from Feb. 13 through Feb. 26, and from Mar. 26 to Apr. 2, 2019. Figure \ref{fig:configuration} shows the photographs and a schematic view of the detector test dewar and optics. The cryogenic blackbody was installed in the detector dewar to serve as a radiation source for measurement of the basic parameters and the system stability. The blackbody was controlled with an accuracy of 1 $m$K by a Lakeshore temperature controller.  The following two apparatuses were also installed in the detector dewar: 1. a pinhole with diameter of $100 \micron$ attached to the blackbody source and 2. a bandpass filter and a ${\rm BaF_{2}}$ window put on the detector mount (Figure \ref{fig:configuration}). The purpose of the former is to illuminate uniform light over the entire detector. Note that the size of the illuminated area was limited by the bandpass filter holder; the diameter of the illuminated area was about 25 mm, corresponding to the clear aperture of the filter mount. The latter restricts the observing wavelength to a narrow band with a center wavelength of 8.6 $\micron$ and a bandwidth of 0.1 $\micron$ because the Si:As IBC detector is sensitive to the wavelength ranging from 2 to 30 $\micron$ \citep[e.g.,][]{2015PASP..127..665R}. 

The array sizes of both the ROIC and SB226 detector are 1024 x 1024 pixels. Both of them have four outputs and do not have reference pixels. The detector array is read out in four channels of alternating columns. An Astronomical Research Cameras Inc. controller (hereafter ARC controller) was used as the warm electronics system to control the detector array. It also amplifies and digitizes the four output signals and then sends this to the data acquisition CPU. A custom data acquisition software developed at NASA Ames Research Center (AIDA) runs on the data acquisition CPU and packages these data into two-dimensional images and sets the detector biases and clocks. The AIDA software stores signal output voltage, which corresponds to the product of the detector output and the gains of the source follower and ARC controller. The file format of the data saved in the data acquisition CPU is FITS and the bit size per pixel is 16. All of the data taken for measurement of the basic parameters of the detector were acquired with uniform time spacing between each pixel sample (sampled "up-the-ramp"). However, data analysis was done with Fowler-4 differences and not slope fits. The pixel clock time for both the ROIC and SB226 ROIC is 10.9 $\mu$s per pixel and it takes about 2.86 s to acquire all 1024 x 1024 pixels in a single read (one frame). There is a reset frame just before beginning each integration. 

\subsection{Preparation \label{subsec:preparation}}
Before measuring the time-series data, the basic detector performance parameters were evaluated from the datasets taken in the first and second cryogenic runs. We measured the following parameter values with an applied detector bias of 1.0 V. The source follower gain is 0.811. The gain of the ARC controller is 9.02. The input capacitance is 33.5 $fF$. The dark current is 11 and less than 1 $e^{-}/s$ at 8 and 7 K, respectively. The read noise is 78 $e^{-}/read$ for correlated double sampling (CDS) and is reduced by the square root of the Fowler pairs. Based on these parameters, the system gain was derived as 28586 $e^{-}$/V. As shown in Table \ref{tab:basic_parameters}, these parameters except for the system gain were very similar to those acquired in the previous evaluation \citep{2003SPIE.4850..890E}. Note that the reason why the system gains measured by the two studies were different is that different detector controllers were in the two studies.

\begin{deluxetable*}{ccc}
\tablecaption{Basic parameters of mid-infrared detector used for this study\label{tab:basic_parameters}}
\tablewidth{0pt}
\tablehead{ \colhead{Item} & \colhead{Measured value (this study)} & \colhead{Value in Ennico et al. (2003)} \\}
\startdata
Applied effective bias & 1.0 V & 0.4 - 3.1 V\\
Source follower gain & 0.811 & 0.823\\
Gain of detector controller & 9.02 &  - \\
Input capacitance at 1 V & 33.5 $fF$ & 33 $fF$ \\
Dark current & $< 1 e^{-}/s$ at 7 K & 0.3 $e^{-}/s$ at 7.5 K\\
Read noise for CDS & 76 $e^{-}/read$ & 73 $e^{-}/read$ at 6 K \\
Read noise for Fowler-4 & 32 $e^{-}/read$ & 40 $e^{-}/read$ at 6 K \\
System gain & 25856 $e^{-}$/V  & 12500 $e^{-}$/V \\
\enddata
\tablecomments{Ennico et al. (2003) used a different detector controller. The gain of the controller was not described in Ennico et al. (2003). The system gain estimated in this study was different from that of the previous study because of using another detector controller.}
\end{deluxetable*}

\begin{figure}[ht!]
\plotone{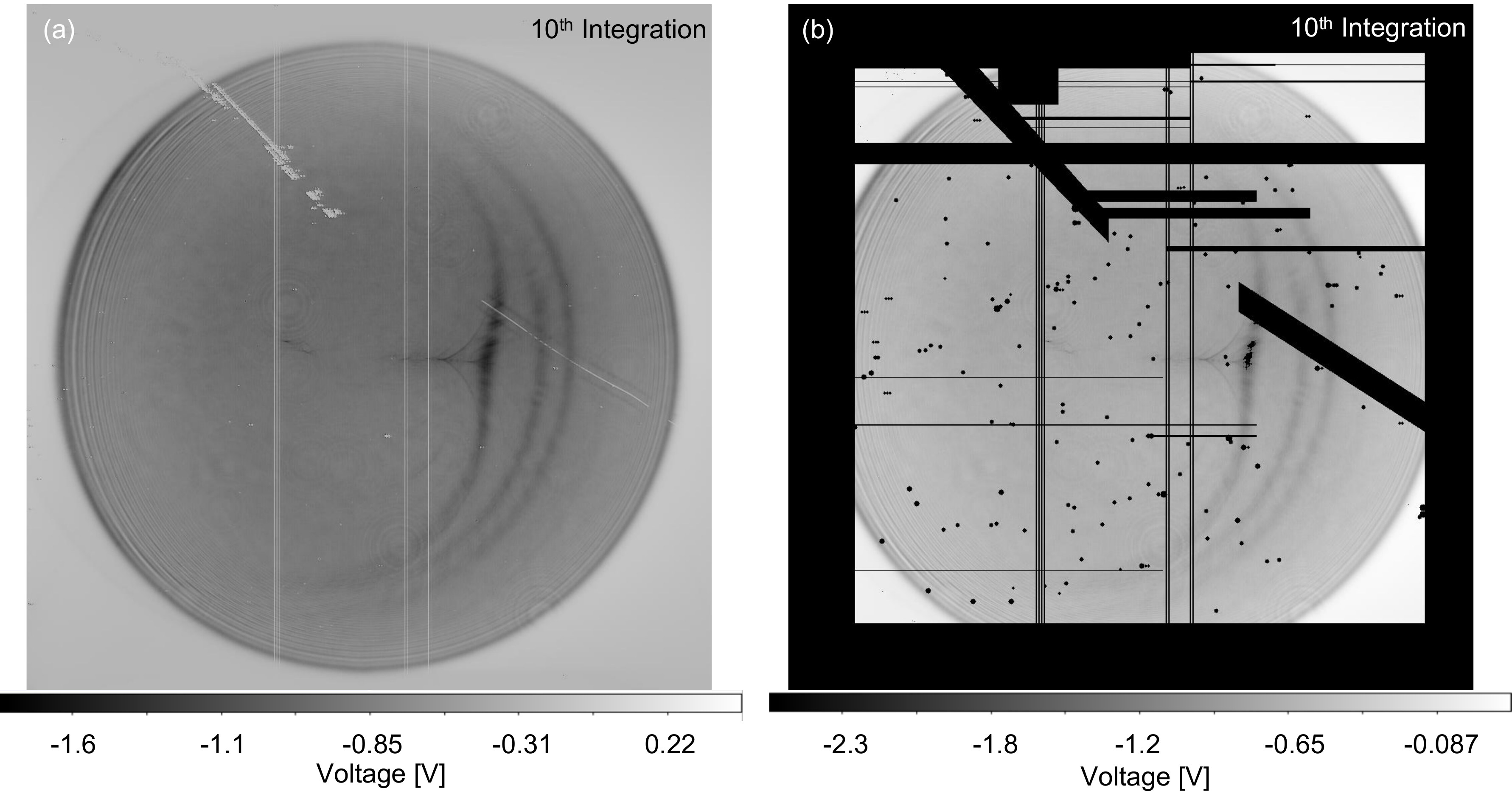}
\caption{ Fowler-4 difference before (a) and after (b) the bad pixel mask was applied. The 10$^{{\rm th}}$ integration of the dataset taken on April 2nd, 2019 was used. The grey scales of the two figures represent the output voltage acquired at the data acquisition CPU. \label{fig:fowler_4}}
\end{figure}

We evaluated the photometric precision of the detector system, using time-series dataset taken on April 2nd, 2019 in the third cryogenic run. The dataset was obtained under the following conditions: 1. the blackbody source turned on and the source temperature was set to 220 K, 2. the detector temperature was set to 7 K and was controlled to a precision of 1 $m$K, 3. each sample-up-the-ramp integration consisted of 10 frames, and 4. the total number of the integrations was 1500. The total elapsed time was 50414 seconds. Note that the standard deviation of the variation of the source temperature over the whole measurement period was 0.46 mK. Therefore, the variation of the light intensity was approximately 46 $ppm$ at 8.6 $\micron$. Because the source temperature randomly changed, the stability better than 5 $ppm$ could be achieved when the number of the coadded integrations is 100; the stability of the detector system could be evaluated with a precision of a few $ppm$. In each integration, a Fowler-4 difference was constructed from the 10 frames, after discarding the first and last frames. A bad pixel mask designed for removing the following regions was applied to each Fowler difference: “Flare” that has additional electrons due to the heater for the detector, regions that do not or weakly respond to the light (i.e., dead pixels), and pixels that have “intermittent noise” generated by the ARC controller and cause anomalies in the time-series data. Figure \ref{fig:fowler_4} shows examples of the Fowler-4 differences before and after applying the bad pixel mask. A total of 30 $\%$ of the detector pixels were masked. Note that the bad pixel mask was also applied to the datasets used for measuring the basic parameters of the detector system. 

\subsection{Gain compensation \label{subsec:gain}}
\begin{figure}[ht!]
\plotone{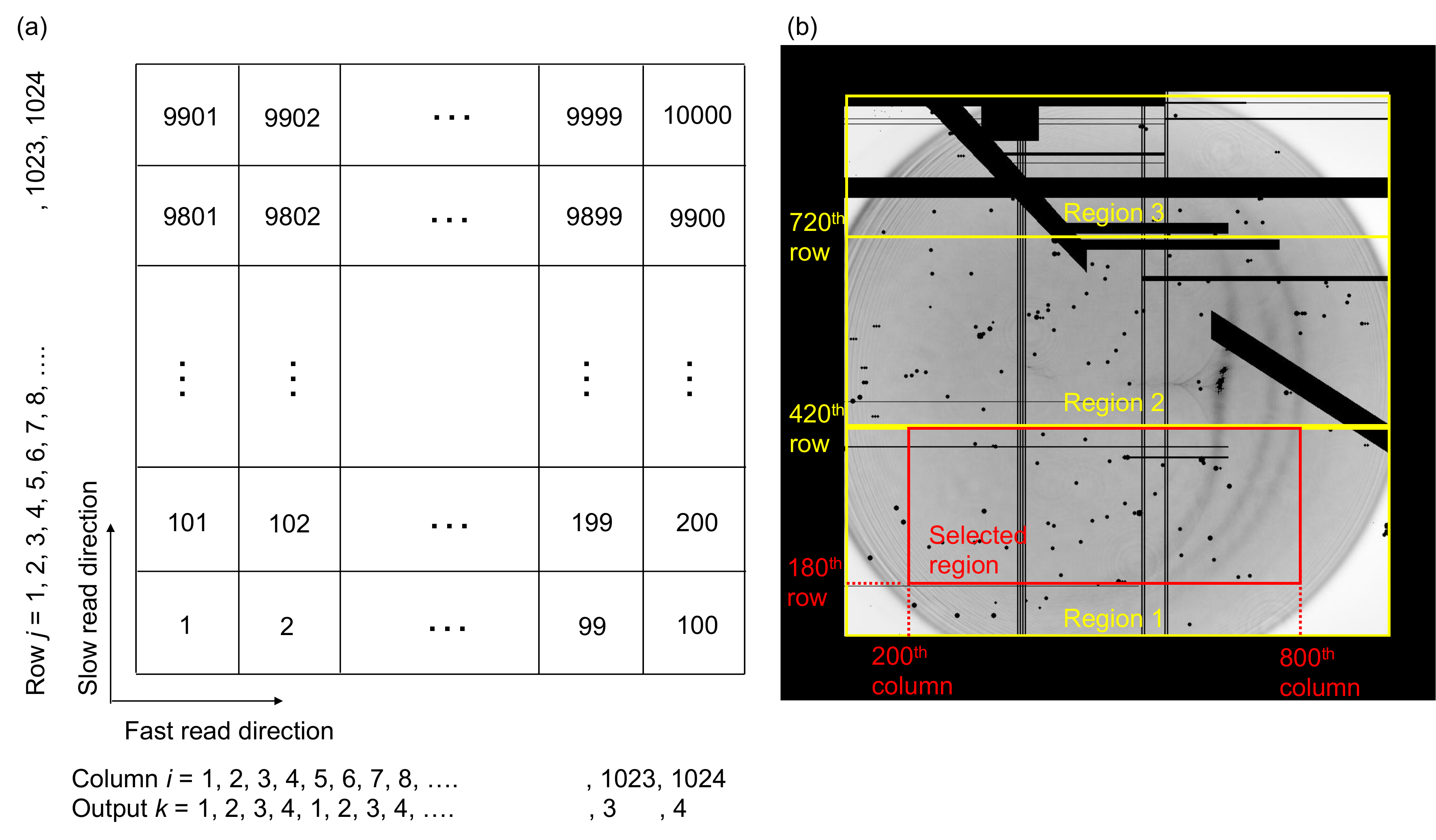}
\caption{ (a) Positions and labels (i.e., numbers) of 10,000 sub-regions. The size of each sub-region was set to 10 x 10 pixels. The horizontal and vertical axes correspond to the fast and slow read directions, respectively. (b) Three regions (indicated by yellow rectangles) divided in terms of the features of the time-series and a region (red rectangle) selected for evaluating the spectro-photometric precision of the detector system. These rectangles were superimposed on the Fowler-4 difference with a bad pixel mask. The yellow and red numbers represent the boundaries of the three regions and selected region, respectively. \label{fig:sub-regions}}
\end{figure}

In this subsection, we describe how the time-series data were constructed. In order to briefly check raw time-series data, we first divided the Fowler-4 difference constructed from each integration into 10,000 sub-regions. Figure \ref{fig:sub-regions}(a) shows the positions and labels (i.e., numbers) for the divided 10,000 sub-regions. Note that 24 rows and 24 columns at the upper and right edges were excluded when the array was divided into 10,000 sub-regions and the excluded pixels are in the masked area shown in Figure \ref{fig:fowler_4} (b); an area of 1000 $\times$ 1000 pixels was used for the analysis. We found that the time-series data of the 10,000 sub-regions can be grouped into three regions along the slow read direction (i.e., along the same column; also see Figure \ref{fig:sub-regions}(b)) in terms of the features of the time-series data. Two features common to the illuminated pixels in Region 1 were observed. 1. The signal continuously decreased by a factor of 0.999 over the first $\sim$ 100 integrations ($\sim$ 56 minutes) and 2. the signal suddenly jumped in the middle of the time-series. The time-series of Region 2 have large noise spikes with amplitudes of 0.1 $\%$ over the entire measurement period, in addition to the first feature (initial signal decrease) of Region 1. Note that the spikes were mainly composed of two values with difference of 0.1 $\%$. The time-series data of Region 3 have only the decreasing signal seen in Regions 1 and 2. Considering that the number of the bad pixels in Region 1 is much smaller than that for Region 3, we selected a region that was fully illuminated by the source from Region 1 (i.e., "selected region" indicated by the red rectangle in Figure \ref{fig:sub-regions}(b)) and then evaluated the stability of the detector system, using the time-series data of their sub-regions.

Here, we found that the signal jump and noise spikes were due to each integration having 3 separate signal gains over the time series, with the pixel location of the gain changes moving slowly in the time series. This is because the slopes of the integrations before and after the signal jump were different. The slopes of the two values that comprise the spikes were also different. Considering that the features of the sub-regions change slowly and relatively uniformly with time, we believe that the signal jumps and large noise spikes were caused by the ARC controller. Because the goal of this study is to measure the photometric precision of the mid-infrared detector with a precision better than 10 $ppm$, we need to compensate these features with an amplitude of 0.1 $\%$.

\begin{figure}
\plotone{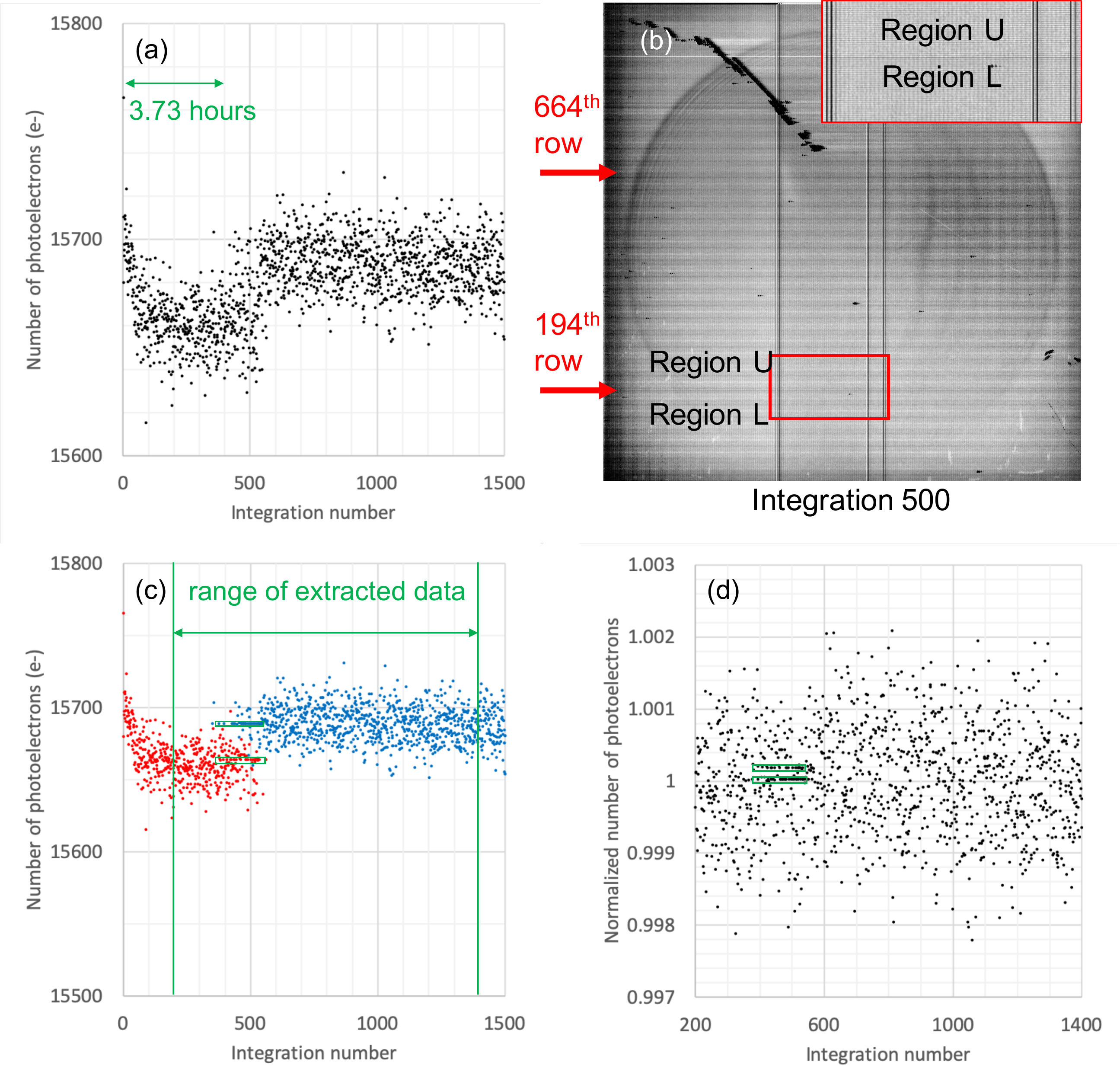}
\caption{(a) Time-series data of sub-region 1821 (within Region 1) indicated by the red rectangle of Figure \ref{fig:sub-regions}(b). The total elapsed time for 400 integrations is approximately 3.73 hours. (b) First frame of the integration number 500. The two red arrows show the two boundaries between three system gains. The voltages of the boundaries in each first frame were much lower than those of the surrounding pixels. Note that, while the two system gains in Region 1 caused the signal change observed in Region 1, the large noise spikes observed in the time-series data of Region 2 were generated due to fluctuation of the boundary position in Region 2. The regions U and L correspond to the regions above and below row 194. The upper right image represents the enlarge view of the region surrounded by the red rectangle. (c) Time-series data divided into two in terms of the two system gains. The time-series of sub-region 1821 was used. The red and blue points show the data belonging to the upper and lower regions of the boundary, respectively. Note that the system gain in Region L was about 0.998 times that of Region U.  The data points surrounded by the two green rectangles contained the gain boundary, and their data values were replaced with the average of the sub-region data segment (red or blue) with the system gain corresponding to the majority of its pixels. (d) Normalized time-series data after the two system gains were compensated. There is no jump of the signal observed in the compensated time-series data. The data points surrounded by the two green rectangles were originally those replaced by the average of each time-series data segment in the panel (c). \label{fig:time-series}}
\end{figure}

Figure \ref{fig:time-series} shows the procedure for constructing the time-series data of a sub-region belonging to Region 1. Figure \ref{fig:time-series}(a) shows the raw time-series data of the sub-region. The signal jumps near the 500$^{\rm th}$ integration were due to each integration having two different signal gains during the 1500 integration sequence, with the location of the gain change moving slowly in different integrations. First, we determined the boundary position of the two system gains in each integration. Here, because the voltage of the first frame for each integration had anomalous values in the row demarking the boundary between the two gain regions, as shown in Figure \ref{fig:time-series}(b), we could correctly determine the position of the gain boundary for each integration. Figure \ref{fig:time-series}(c) shows the time-series data divided into two segments with different system gains (red and blue). We masked the data values for when the gain boundary was within sub-region 1821 and replaced them with the average for the data segment with the gain corresponding to the majority of the sub-region's pixels. The masked and replaced data points are surrounded by the green rectangles of Figure \ref{fig:time-series}(c). We masked and replaced approximately 60 of the 1200 total integrations for each sub-region in this manner. Next, considering that the ratio of the two system gains are unknown and the blackbody source is stable with a precision of a few $ppm$, we compensated the system gains such that the time-series data is continuously connected without any jumps after the system gains are compensated. Specifically, the ratio of the two system gains was derived under the condition that the averages of 50 integrations before and after the jumps of the signal should be equal. We scaled one of the two time-series data segments by this ratio so that the means of both regions were identical. We also extracted the data for evaluating the stability from the whole time-series data (i.e., 1500 integrations) such that the number of the integrations is multiple of 400 for each time-series data. This is because the total elapsed time of the 400 integrations is 3.73 hours, about 3 times the transit duration of a habitable planet orbiting a late M-type star (see Section \ref{subsec:models}). We removed integrations 1 to 200 and 1400 to 1500 because the signal continuously decreased in the former period. We normalized the extracted time-series data to 1. Figure \ref{fig:time-series}(d) shows the normalized time-series data after the system gains were compensated. Finally, the extracted time-series data was divided into three datasets such that the number of the integrations for each dataset is 400. A co-added dataset consisting of 400 integrations was constructed to maximize the detected signal over the period of an exoplanet transit.

Note that we also observed large noise spikes in the time-series data of Region 2 in the stability test of the bare ROIC. Therefore, the two separate signal gains are thought to be caused by our ARC controller; the above gain compensation would be not necessary for actual observations with a space-based observatory that had more stable detector control electronics.
     
\section{Results} \label{sec:results}
In this section, we evaluate the photometric stability of the JWST MIRI prototype Si:As IBC detector, using the time-series data constructed in Section \ref{subsec:gain}. After we show the measured photometric precision of the mid-infrared detector in Section \ref{subsec:photometric},  we discuss the correlation observed in the time-series data and its impact on the photometric precision in Section \ref{subsec:correlation}. Given that a part of the detector is exposed by a stable internal source, we compensate the correlated gain variation with the reference light constructed from the internal source in Section \ref{subsec:compensation}. We also investigate the cause of the correlation from the point of view of the effective detector bias in Section \ref{subsec:effective_bias}.  

\subsection{Photometric precision \label{subsec:photometric}}
\begin{figure}
\plotone{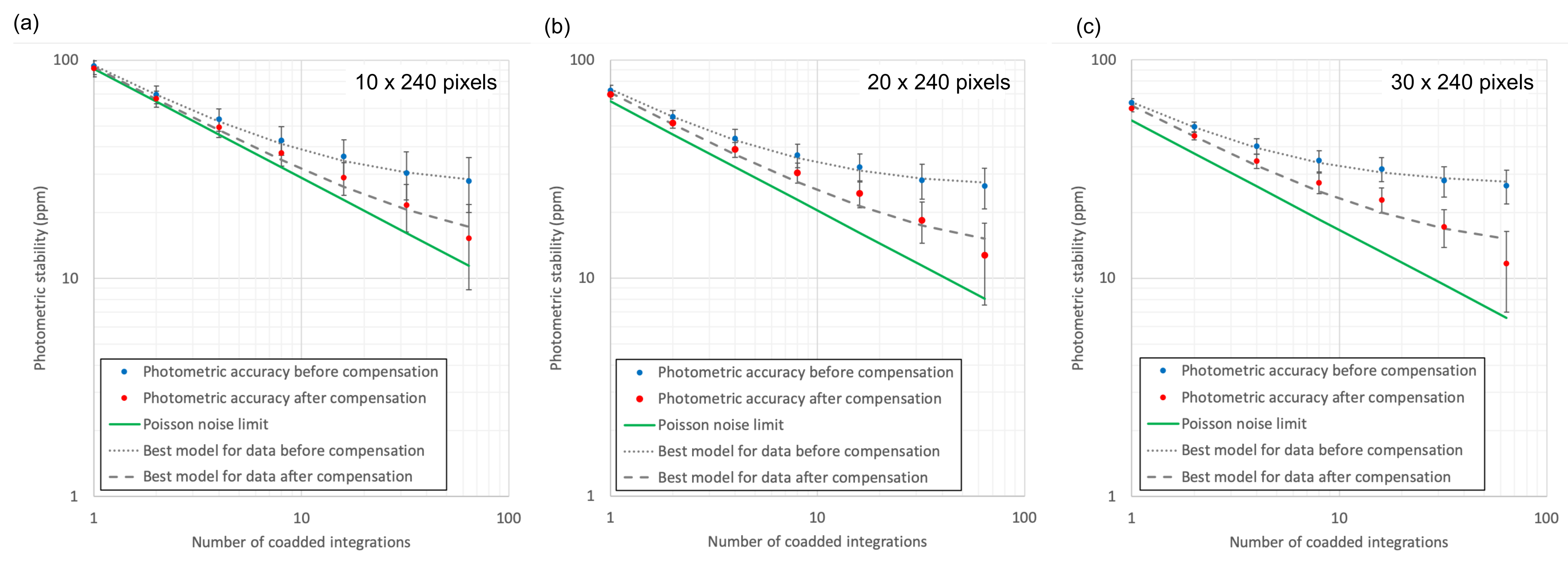}
\caption{Photometric precision of three boxes with sizes of (a) 10 x 240, (b) 20 x 240, and (c) 30 x 240 pixels, as a function of the number of the co-added integrations. The time-series of each box was generated by co-adding the time-series data of several 10 x 10 pixels sub-regions. The blue and red dots show the averages of the photometric precision over all the (a) 60, (b) 30, and (c) 20 boxes within the selected region (indicated by the red rectangles in Figure \ref{fig:sub-regions}(b)) before and after the gain compensation, respectively. The common system gain was compensated with reference time-series data constructed from all the pixels within the green rectangle shown in Figure \ref{fig:correlation_model11_no_coadd}(a). The green line is the random noise limit for each box under the condition that the number of the photoelectrons per pixel per integration is 18000 $e^{-}$. The grey dotted and dashed lines represent the best models for the random and systematic noises included in the time-series data before and after the common variation was compensated, respectively. The systematic noise before and after the gain compensation were 26.3 and 12.8 $ppm$, respectively. The best model for the random and systematic noises was derived through Monte Carlo simulation such that the difference between the noise model constructed from Equation (3) and the measured values is minimized. The measurement values were assumed to follow a Gaussian distribution with FWHM of 1-sigma measurement uncertainty. The measurement uncertainty indicated by the error bar is the standard deviation of the photometric precision for all the boxes within the red rectangle shown in Figure \ref{fig:sub-regions}. \label{fig:photometry}}
\end{figure}

We evaluated the photometric precision of the mid-infrared detector, using the constructed time-series data of the sub-regions with a size of 10 x 10 pixels. Note that, in this paper, the photometric precision was defined as 
\begin{equation}
\sigma_{Photometry} = \frac{\sigma_{data}}{<S>_{data}},
\end{equation} 
where $\sigma_{data}$ and $<S>_{data}$ are the standard deviation and average of the time-series data, respectively. In order to evaluate the photometric precision with photon-noise-limited value of 10 $ppm$, we grouped the time-series data of some of the sub-regions within the red rectangle shown in Figure \ref{fig:sub-regions}(b) in three ways and also co-added the integrations. We call the spatially co-added region "box." Note that the dimensions of the boxes combined in three ways were 1. 10 x 240 pixels, 2. 20 x 240 pixels, and 3. 30 x 240 pixels along the slow and fast detector read directions, respectively. The numbers of the boxes within the selected region (indicated by the red rectangle shown in Figure \ref{fig:sub-regions}) are 1. 60, 2. 30, and 3. 20, respectively. The blue dots shown in Figure \ref{fig:photometry} show the means of the photometric precision over the 60, 30, and 20 boxes as a function of the number of the co-added integrations. We found that the photometric precision was approximately 26.3 $ppm$ at high co-added signal levels for the three boxes with different sizes. In addition, the achieved precision was worse than the ideal performance that is limited by the fluctuation of the random noise (hereafter random noise limit). As discussed in Section \ref{subsec:compensation}, the gain variation common to the boxes limited the photometric variation.

The random noise limit of each pixel for each integration, $\sigma_{random, pixel, exposure}$, was calculated as follows:
\begin{equation}
\sigma_{random, pixel, exposure} = \frac{\sqrt{Q_{ph}+Q_{dark}+Q^{2}_{read}}}{Q_{ph}+Q_{dark}},
\end{equation} 
where $Q_{ph}$ is number of the photoelectrons per a pixel per an integration, $Q_{dark}$ is dark current per an integration, and $Q_{read}$ is read noise of the detector system per integration. The random noise limit is improved by the square root of the number of pixels per a box and the number of the co-added integrations. Given that the number of the photoelectrons accumulated to each pixel at the end of each integration is $\sim$ 18000 $e^{-}$, the random noise limit is 91.3 $ppm$ for the box with a size of 10 x 240 pixels. Note that, because the time-series of each 10 x 10 pixels sub-region was constructed by co-adding three time-series data (see Section \ref{subsec:compensation}), the random noise limit was improved by a factor of $\sqrt{3}$. The read noise for each pixel was set to 39 $e^{-}$/integration for the Fowler-4 pairs; the read noise was insignificant compared to the Poission noise. Because the dark current is less than 1 $e^{-}/s$ at the detector temperature of 7K, we ignored the dark current in this calculation.

\subsection{Correlation of time-series data \label{subsec:correlation}}
\begin{figure}
\plotone{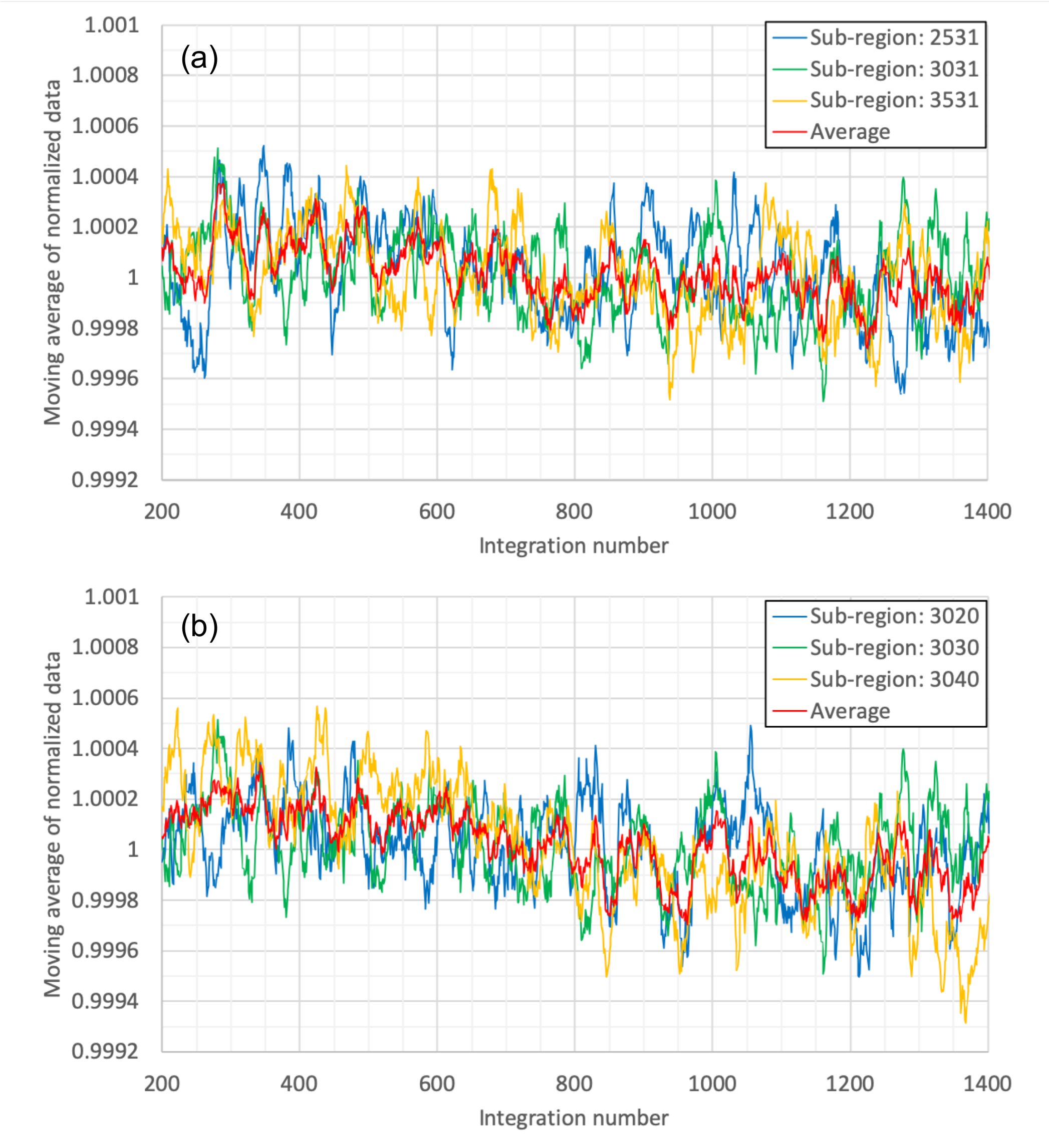}
\caption{ 20-integration moving average of the normalized time-series data of three 10 x 10 pixel sub-regions (blue, green, and yellow lines). The three sub-regions are aligned along the slow (a) and fast (b) read directions. The red line in each panel represents the average of the time-series data of the three sub-regions. \label{fig:correlation_subregions}}
\end{figure}

We discuss the impact of the following two features observed in the time-series data on the photometric precision. 1. There are no correlations in the time-series data of different sub-regions with a size of 10 x 10 pixel at the few hundred ppm level, and 2. there is a common variation observed in the time-series data of the boxes. Figure \ref{fig:correlation_subregions} shows a 20-integrations moving average of the normalized time-series data of three 10 x 10 pixel sub-regions. The dispersion of the three combined sub-regions (indicated by red line in Figure \ref{fig:correlation_subregions}) was smaller than that of each time-series; the unique variation accompanying with the time-series data of the sub-regions was smoothed out through averaging each single sub-region. In other words, the photometric precision of a box could be more improved as the number of pixels per a box increases. 

\begin{figure}
\plotone{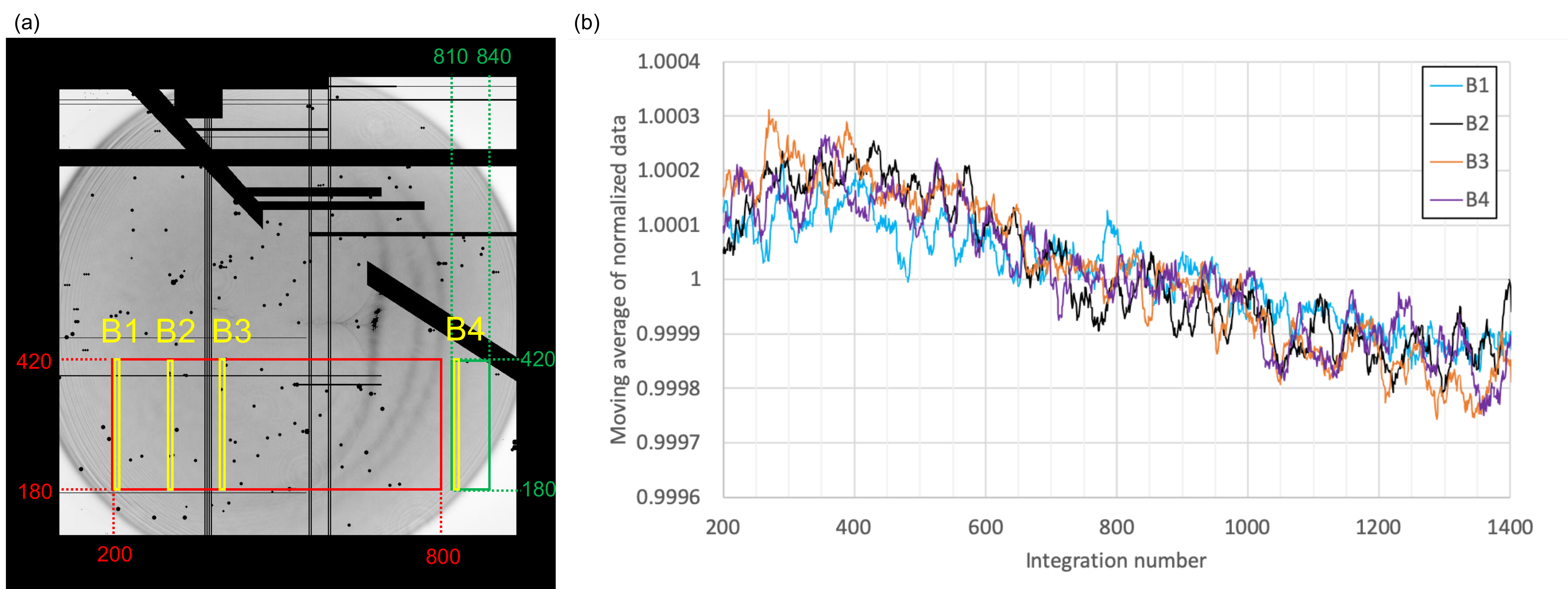}
\caption{(a) Positions of four boxes with 10 x 240 pixels (B1, B2, B3, and B4) used for evaluating the correlation of their time-series data. The boxes are numbered in yellow. The green rectangle shows the region used for constructing a reference time-series data. Note that, although a part of the region surrounded by the green rectangle was not exposed, all of the pixels within the green rectangle were used for constructing the reference curve because the values of the pixels that were not illuminated by the blackbody source is much lower than those of the illuminated pixels. (b) 20-integrations moving average of the time-series data of the selected four boxes. \label{fig:correlation_model11_no_coadd}}
\end{figure}

On the other hand, when the number of pixels per the box is more than 2000, a correlation between the time-series of the boxes was observed. Unlike the time-series data of the 10 x 10 pixels sub-regions, there is a correlation at a 100 ppm level between the three different boxes (see Figure \ref{fig:correlation_model11_no_coadd}(b)). The time-series data continuously decreased with a rate of 100 $ppm$ per hour. As a result, the photometric precision was approximately 26.3 $ppm$, which was worse by a factor of 3 than the random noise limit. 

Although a slow signal drift like this has not been observed in data from the similar Spitzer/IRAC Si:As IBC detectors \citep{2009ApJ...703..769K, 2012ApJ...754...22K} and the Wide-field camera 3 mounted on HST\citep{2014Natur.505...69K},  we do not understand whether this common gain drift originates in the detector or our older generation ARC controller at this point.

\subsection{Compensation of common gain variation \label{subsec:compensation}}
The common gain variation to the time-series data of the boxes limits the photometric precision. However, focusing on a fact that the variation observed at the high co-added signal levels is common over the selected region (indicated by the red rectangle shown in Figure \ref{fig:sub-regions}(b)), we found that the common variation can be compensated with reference light generated by a stable internal blackbody source. Assuming that the green rectangle shown in Figure \ref{fig:correlation_model11_no_coadd}(a) is exposed by an internal blackbody source, the time-series data of each box was compensated with the reference time-series data constructed from the region indicated by the green rectangle. Note that all pixels in each integration of each box were scaled by the 20-integration (10.5 minutes) moving average signal of the reference area, normalized to its initial value. The red dots shown in Figure \ref{fig:photometry} represent the compensated photometric precision of the time-series data.  

In order to quantitatively investigate the effect of the compensation on the photometric precision, we evaluated the excess noise component generated due to the detector system before and after the gain compensation. The excess noise is thought to be composed of random and systematic components. While the random component is reduced by averaging a number of the pixels and binning the integrations, the systematic noise does not depend on the number of pixels per a box and the number of the coadded integrations. Therefore, the total noise, $\sigma_{total}$, can be written as 
\begin{equation}
\sigma_{total}(N_{pixel}, N_{coadd}) = \sqrt{\sigma^2_{total, random}(N_{pixel}, N_{coadd})+\sigma^2_{total, systematic}},
\end{equation}
where $N_{pixel}$ is the number of pixels per a box, $N_{coadd}$ is the number of the co-added integrations, and the total random noise is composed of the random noise determined by the Poisson noise and read noise, $\sigma_{random}$, and an unknown random component generated due to the system gain variation, $\sigma_{gain}$:
\begin{equation}
\sigma_{total, random}(N_{pixel}, N_{coadd}) = \sqrt{\sigma^2_{random}(N_{pixel}, N_{coadd})+\sigma^2_{gain}(N_{pixel}, N_{coadd})}.
\end{equation}
Based on these considerations, we estimated the random and systematic noises included in the time-series data through Monte Carlo simulation. The best model for the random and systematic noises was derived such that the difference between the noise model of Equation (3) and the measured values is minimized. As shown in Figure \ref{fig:photometry} (indicated by grey dotted line), the best model has a good agreement with all the models. While the compensation did not have an impact on the random component, the systematic noise could be reduced by a factor of 2 after the gain compensation; the systematic noise before and after the compensation was approximately 26.3 and 12.8 $ppm$ for the three boxes with different sizes, respectively. We also found that the measured precision improved with time was consistent with Equation (3) and no other significant random noise term was required (See Figure \ref{fig:photometry}). 

\subsection{Effective bias \label{subsec:effective_bias}}
We operated the mid-infrared Si:As IBC detector at the effective bias of 1 V because the basic parameters of the detector were measured at that voltage in a previous study \citep{2003SPIE.4850..890E}. However, because the IR-active layer in this type of detector is not fully depleted unless the effective bias is more than 1.5V \citep{2015PASP..127..665R}, these detectors are normally operated at higher effective bias. The thick depletion layer leads to enhance the well capacity as well as the quantum efficiency \citep{2015PASP..127..665R}. For example, while the well capacity of our detector was approximately 50,000 $e^{-}$ at 1 V applied detector bias \citep{2003SPIE.4850..890E}, that for the focal plane arrays of the Mid-Infrared Instrument (MIRI) mounted on JWST is 250,000 $e^{-}$ at 2.2 V and the system gain is approximately 210,000 $e^{-}$/V \citep{2015PASP..127..675R}. 

The lower detective quantum efficiency (DQE) and lower well capacitance are achieved at lower effective detector biases. Our use of 1V detector bias may have resulted in variation of the depletion region depth during each integration and achieving less than optimum DQE for this device. However, these effects should 
be the same for all integrations, so they should not impact the measured photometric precision.  Higher detector bias levels should be considered when using similar detectors in real astronomical instruments in order to maximize DQE and to minimize latent images \citep{2015PASP..127..665R}. 

\section{Application to transit spectroscopy} \label{sec:discussion}

In this section, we discuss an application of the measured photometric precision to transit spectroscopy. We assume that a densified-pupil spectrograph will be used for exoplanet transit observations and we  introduce models for estimating the spectro-photometric precision in Section \ref{subsec:models}. We show the expected spectro-photometric precision of the mid-infrared detector in Section \ref{subsec:spectrophotometric_variation}. Note that, because the densified pupil spectrograph largely reduces the other systematic noises such as image motion on the intra- and inter-pixel variation compared to general spectrographs, its spectro-photometric precision would be close to the expected one translated from the time-series data of the detector itself.

\subsection{Models \label{subsec:models}}
\begin{figure}
\plotone{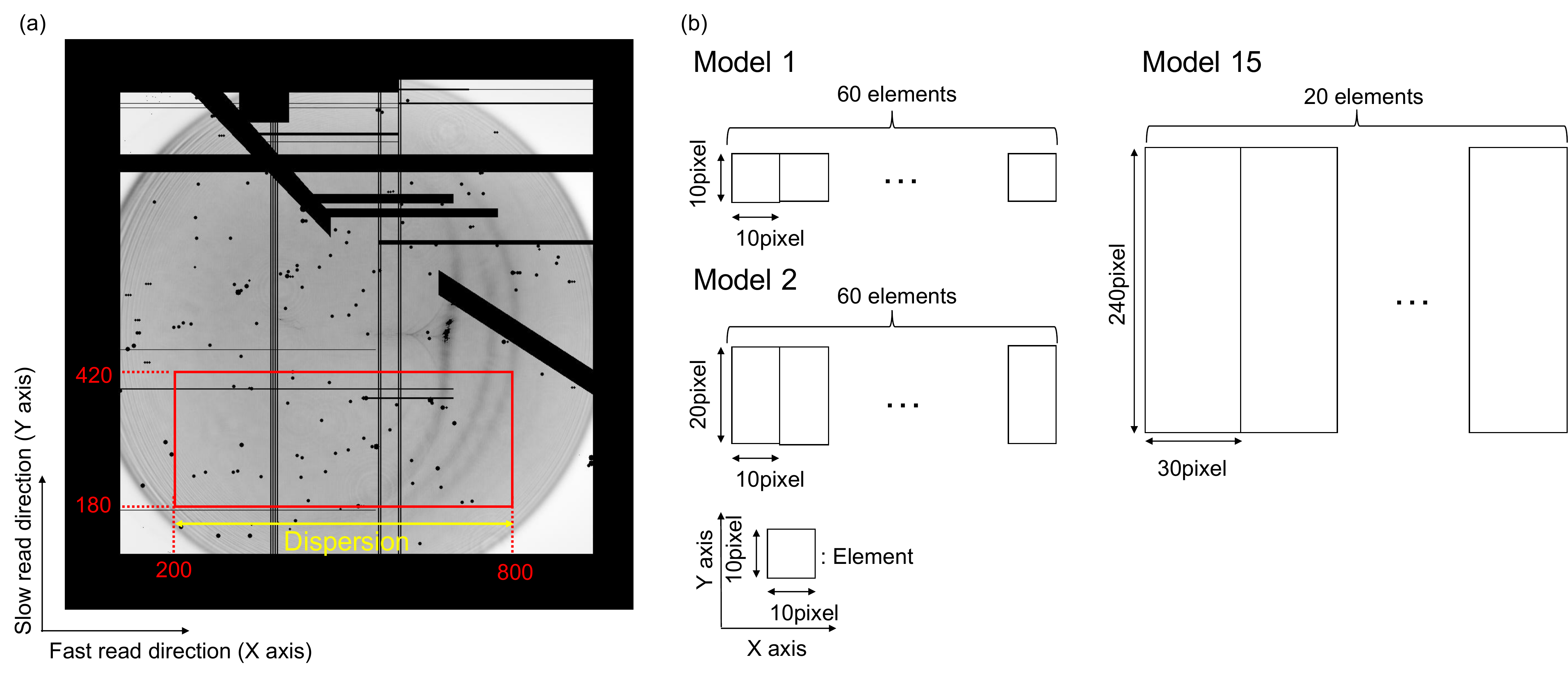}
\caption{(a) Direction of dispersion determined for simulating the 16 models compiled in Table \ref{tab:models}. The red rectangle shows a detector region, in which the spectral resolution elements of each model were selected. The dispersion direction corresponds to the fast detector read direction. The 16 models were constructed from the time-series data of the 10 x 10 pixels sub-arrays within the red rectangle. The X and Y axes correspond to the fast and slow read directions, respectively. (b) Schematic diagrams of the models M1, M2, and M15. \label{fig:models}}
\end{figure}

We modeled the densified pupil spectrograph to translate the time-series data taken under the flood illumination to the expected spectro-photometric data when used in transit observations. Here, the densified pupil spectrograph can freely spread the object light over the entire detector and can change the number of pixels per a spectral resolution element (hereafter element). In order to investigate the dependency of the number of pixels per an element on the translated spectro-photometric precision, we prepared 16 models for the densified pupil spectrograph. Each model has different number of pixels per an element. We set the minimum number of pixels per an element to 100 because this was the size of the sub-regions used in our detector photometric precision analysis and gain correction technique. First, the region selected in Section \ref{subsec:gain} (i.e., red rectangle shown in Figure \ref{fig:models}(a)) was divided into multiple elements. In order to estimate the expected spectro-photometric precision, the direction of the dispersion of the densified pupil spectrograph was defined to be in the fast read direction (along a row in Figure \ref{fig:models}(a)). The elements of each model were aligned along the dispersion direction, as shown in Figure \ref{fig:models}(b). Table \ref{tab:models} compiles the parameters of each model. 

\begin{figure}
\plotone{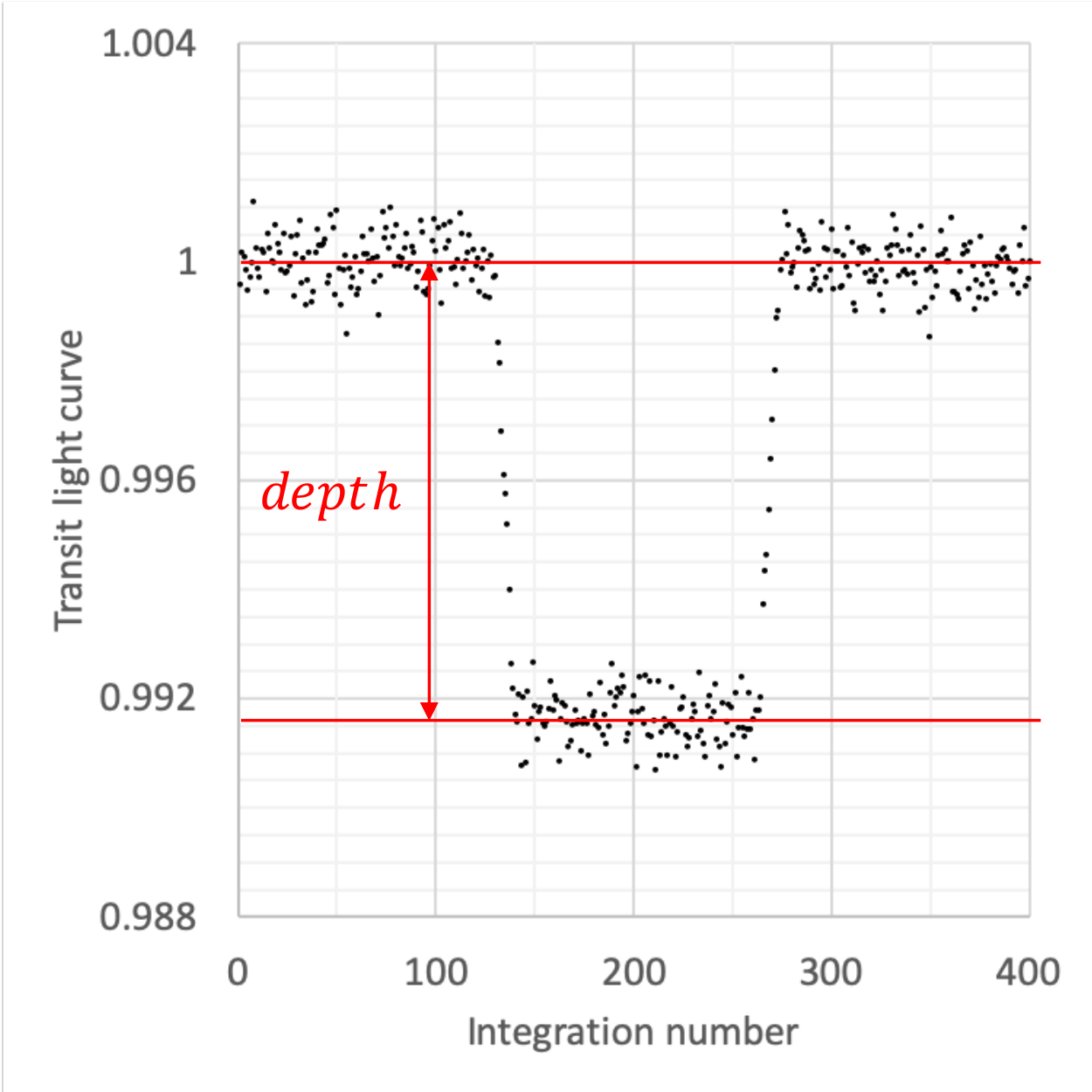}
\caption{Simulated transit light curve composed of 400 integrations for the model M1. An ideal transit light curve was injected into the time-series data of sub-region 1821. Given that a temperate transiting planet with radius of 1 $R_{\oplus}$ orbits a late-M type star, the ideal transit light curve was calculated based on a previous study \citep{2002ApJ...580L.171M}. Note that the elapsed time of 400 integrations are almost corresponding to three times the transit duration of the assumed planet. The depth was estimated as the difference between the means of the light curve in the baseline and in-transit eclipse. \label{fig:light_curve}}
\end{figure}

Next, we simulated the transit observation by injecting an ideal transit light curve to the time-series data constructed in Section \ref{subsec:gain}. Based on the analytic light curve \citep{2002ApJ...580L.171M}, we modeled the ideal transit light curve for a transiting planet orbting late M-type star with effective temperature of 2500 K and radius of 0.1 $R_{\odot}$. The transiting planet is a terrestrial planet with effective temperature of 288 K and radius of 1 $R_{\oplus}$. Given that the albedo of the target is 0.306, the semi-major axis of the planet was set to 0.015 AU. The inclination was set to 90 degrees. Because the effect of the limb darkening on the transit light curve is very small at the mid-infrared wavelengths, no limb darkening was assumed to exist in the ideal transit light curve. The transit depth is approximately 0.0084. Figure \ref{fig:light_curve} shows an example of the simulated transit light curve composed of 400 integrations.

\begin{deluxetable*}{cccccc}
\tablecaption{Models and expected spectro-photometric precision\label{tab:models}}
\tablewidth{0pt}
\tablehead{ \colhead{Model} & \colhead{Num. of elements} & \colhead{Num. of pixels per element} &\colhead{Size of element} &  \colhead{Random noise limit} & \colhead{Spectro-photometric precision}\\
 \colhead{} & \colhead{} & \colhead{pixels} & \colhead{pixels} & \colhead{ppm} & \colhead{ppm}}
\startdata
M1 & 60 & 100 & 10 x 10 & 49.3 & (3.2, 61.8)\\
M2 & 60 & 200 & 20 x 10 & 34.8 & (3.6, 42.6)\\
M3 & 60 & 300 & 30 x 10 & 28.4 & (7.3, 36.9)\\
M4 & 60 & 400 & 40 x 10 & 24.6 & (6.4, 32.4)\\
M5 & 60 & 600 & 60 x 10 & 20.1 & (9.3, 28.7)\\
M6 & 60 & 800 & 80 x 10 & 17.4 & (13.4, 25.8)\\
M7 & 60 & 1000 & 100 x 10 & 15.6 & (14.3, 22.7)\\
M8 & 60 & 1200 & 120 x 10 & 14.2 & (12.6, 19.4)\\
M9 & 60 & 1600 & 160 x 10 & 12.3 & (7.8, 15.6)\\
M10 & 60 & 2000 & 200 x 10 & 11.0 & (8.5, 13.4)\\
M11 & 60 & 2400 & 240 x 10 & 10.1 & (8.7, 11.8)\\
M12 & 30 & 3600 & 180  x 20 & 8.2 & (7.8, 10.8)\\
M13 & 30 & 4800 & 240 x 20 & 7.1 & (9.3, 9.4)\\
M14 & 20 & 6000 & 200 x 30 & 6.4 & (9.2, 8.2)\\
M15 & 20 & 7200 & 240 x 30 & 5.8 & (9.3, 7.5)\\
M16 & 15 & 9600 & 240 x 40 & 5.0 & (8.7, 7.0)\\
\enddata
\tablecomments{The size of the element is X x Y pixels, where the X and Y axes correspond to the fast and slow read directions, consistent with Figure 8. 
The expected spectro-photometric precision column gives the (absolute, relative) values, defined as the mean and standard deviation of the measurement errors for all the elements
belonging to each model, respectively.}
\end{deluxetable*}

\subsection{Expected spectro-photometric precision \label{subsec:spectrophotometric_variation}}

We assess our ability to recover the transit spectrum by measuring the difference between the injected and recovered transit signal for each model. We define the absolute spectro-photometric precision as the error averaged over all spectral elements and the relative spectro-photometric precision as the standard deviation of the different spectral elements, with both expressed as fractions of their respective signals in ppm. The absolute precision is important for estimating the absolute depths of atmospheric features through dividing out the host star, and the relative precision is important for determining the relatives strengths of different spectral features.

The procedure for estimation of the spectro-photometric precision is as follows. We first defined the measurement error of transit depth as the difference between the measured transit depth and the injected one, 0.0084, and then estimated the error for each element of the light curve. Note that, as shown in Figure \ref{fig:light_curve}, the measurement of the transit depth was calculated as the difference between the means of the baseline and in-transit eclipse. We estimated the measurement errors of transit depth for all of the elements in each model. 

\begin{figure}
\plotone{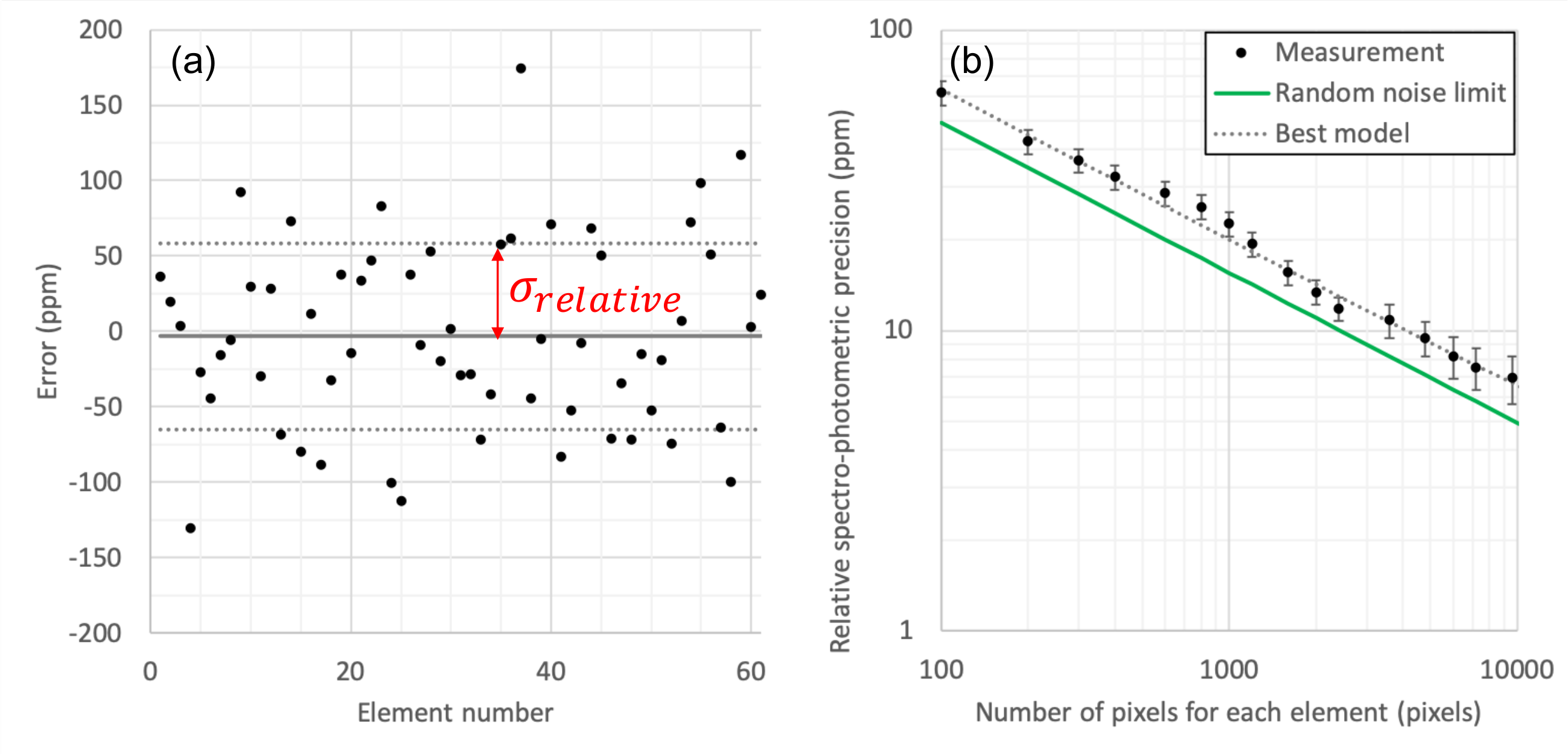}
\caption{(a) Measurement errors of transit depth for the elements of the model M1. The grey solid and dotted lines represent the mean and standard deviation of the measurement errors for all the 61 elements, respectively. The standard deviation indicated by red $\sigma_{relative}$ corresponds to the relative spectro-photometric precision for the model M1. (b) Relative spectro-photometric precision as a function of the number of pixels per a resolution element. Each data point represents the relative spectro-photometric precision of each model. The green solid line shows the random noise limit and the grey dotted line is the best model for the random and systematic noises included in the simulated spectrum. Assuming that the number of the photoelectrons per pixel per integration is 18000 $e^{-}$, the random noise limit was calculated. The best model was derived through Monte Carlo simulation such that the difference between the model constructed from Equation (3) and the measurement result is minimized. The measurement values were assumed to follow a Gaussian distribution with FWHM of 1-sigma measurement uncertainty. The uncertainty of each measurement, $\sigma_{measure}$, was calculated based on $\sigma_{measure} = \frac{\sigma_{relative}}{\sqrt{2(N-1)}}$, where $N$ is the number of the resolution elements for each bin. \label{fig:spectro-photometry}}
\end{figure}

We next calculated the average and standard deviation of the measurement errors of transit depth for each model. Figure \ref{fig:spectro-photometry}(a) shows the measurement errors of transit depth for all the elements belonging to the model M1. The Model 1 synthetic data set was measured to have a mean transit depth error of 3.2 $ppm$ (offset from the injected depth) and a standard deviation of 61.8 $ppm$. In other words, the absolute and relative expected spectro-photometric precision for the model M1 is 3.2 and 61.8 $ppm$, respectively. In the same way, we also derived the absolute and relative precision of the other models (see Table \ref{tab:models}) and then investigated the relation between the precision and the number of pixels per an element (see Figure \ref{fig:spectro-photometry}(b)). We found that, while the absolute precision is almost constant ($\sim$ 10 $ppm$) over the 16 models, the relative spectro-photometric value was almost reduced by the square root of the number of pixels per an element and was approximately 7 $ppm$ at the highest co-added signal level. In addition, the measured precision is larger by a factor of only 1.3 than the random noise limit. The reason why the relative spectro-photometric precision was better than the photometric precision is that the common signal variation, which limited the photometric precision, does not degrade the relative precision. Focusing on a fact that the relative precision is the standard deviation of the measurement errors of transit depth for all the elements belonging to each model, the relative level is independent of the common variation observed in the time-series data; only the residuals from the common variation in each time-series data give an impact on the relative spectro-photometric precision. 

Note that the random noise limit for the spectro-photometric observation was calculated as follows. Because the transit depth is determined by the difference between the baseline and in-transit eclipse, its measurement error, $\sigma_{depth}$, is 
\begin{equation}
\sigma_{depth} = \sqrt{\sigma^{2}_{baseline}+\sigma^{2}_{eclipse}},
\end{equation} 
where $\sigma_{baseline}$ and $\sigma_{eclipse}$ are the measurement errors of the baseline and in-transit eclipse, respectively. Therefore, the random noise limit for the measurement of transit depth can be calculated from $\sigma_{baseline}$ and $\sigma_{eclipse}$ derived according to Equation (2). When the number of the photoelectrons per pixel per integration is 18000 $e^{-}$, the random noise limit of the spectro-photometric observation for the model M1 is 47.6 $ppm$. The numbers of the integrations for each baseline and in-transit eclipse are 127 and 122, respectively. We note that, as discussed in Section \ref{subsec:compensation}, because the time-series data used for estimating the spectro-photometric precision was constructed by co-adding three time-series data, the random noise limit for the observation was improved by a factor of $\sqrt{3}$. The random noise error for each model is compiled in Table \ref{tab:models}.

Finally, we evaluated excess noise generated by the detector warm electronics system in the simulated transit spectra. In the same way as the calculation of the random and systematic noises attaching to the photometric measurement in Section \ref{subsec:photometric}, we estimated the random and systematic noises included in the simulated transit spectrum through Monte Carlo simulation. As shown in the grey dotted line in Figure \ref{fig:spectro-photometry}(b), the relative spectro-photometric precision of all the models could be explained by the best model with the random and systematic noises of 63.2 and 1.7 $ppm$. The random noise is higher than the systematic component even at high co-added signal levels. Note that the data points were assumed to follow a Gaussian distribution with FWHM of 1-sigma measurement uncertainty indicated by vertical error bars. 

Thus, increasing the number of pixels per an element is useful for improving the relative spectro-photometric precision. As discussed in Section \ref{subsec:models}, because the densified pupil spectrograph spreads the light over the entire detector with multiple sub-pupil spectra \citep{2016ApJ...823..139M}, the densified pupil spectrograph would be helpful for high precision transit spectroscopy.

\section{Conclusion} \label{sec:conclusion}
In order to measure the photometric precision of a JWST MIRI prototype mid-infrared Si:As IBC detector, we acquired its long-term time-series data taken under the ideal conditions that the entire detector was uniformly illuminated by a highly stable cryogenic source (the temperature was controlled with an accuracy of 1 $m$K). This measurement was conducted over periods of $\sim$ 10 hours. We prepared the time-series data for estimating the photometric precision of the detector as follows; 1. a region for analyzing the time-series data was selected, 2. the selected region was divided into multiple sub-regions with a size of 10 x 10 pixels, 3. the jump of the signal observed in the time-series data of the selected sub-regions was compensated, 4. 1200 integrations were selected from the whole time-series data (i.e., 1500 integrations) and were equally divided into three, and 5. a dataset composed of 400 integrations for each sub-region was constructed through averaging the three divided sub-datasets. Note that the total elapsed time of the 400 integrations is 3.73 hours, corresponding to at least three times the transit duration of a habitable planet orbiting a late M-type star. After these procedures, we estimated the photometric precision of the detector as the standard deviation of the time-series data. We found that, at high co-added signal levels, the photometric precision was 26.3 $ppm$, almost corresponding to three times the random noise limit determined by the Poisson noise and readout noise.

There were two features observed in the constructed time-series data. One is no correlation at the few hundred ppm level between the 10 x 10 pixel sub-regions and another is a correlation at one hundred $ppm$ level among boxes with sizes more than 2000 pixels. The correlation was the long-term variation that continuously decreased with a rate of 100 $ppm$ per hour. Thanks to the former feature, the photometric variation could be improved by co-adding pixels and integrations. In contrast, as the number of the co-added pixels and integrations increases, the photometric variation was gradually limited because of the latter correlation. However, focusing on the fact that the variation is common to all detector pixels, we determined that the long-term variation could be compensated with a reference time-series data that was constructed by averaging the time-series data of the pixels within a different region from the selected one. The photometric precision could be improved by a factor of at least 2, 12.8 $ppm$. At this point we do not understand whether this common gain drift originates in the detector or our older generation ARC controller, so we cannot determine whether it would impact a spaceflight system without a stable flux reference source.

We also translated the measured photometric precision to the expected spectro-photometric precision. We simulated the densified pupil spectrum, injecting an ideal transit light curve to the constructed time-series data. We found that, while the absolute expected spectro-photometric precision is almost constant, the relative precision could be almost improved by the square root of the number of pixels per a resolution element. At the high co-added signal levels, the total noise could be reduced down to 7 $ppm$ and the systematic noise was 1.7 $ppm$. The reason why the relative precision was better than the photometric one is that the relative precision was independent of the common photometric variation, which limited the photometric precision.

Based on these measurements, increasing the number of pixels per a resolution element is important for improving the relative spectro-photometric precision. Higher stability warm detector electronics are required for achieving high photometric precision. We also have to emphasize that it is important for transit spectroscopy to reduce the other instrument systematic errors such as the image movement on intra- and inter-pixel variation and image motion loss on the field stop and slit. Thus, the densified pupil spectrograph that spreads spectra of pupil segments over the entire detector is thought to have an important role in high-precision transit spectroscopy.

We are currently developing a mid-IR cryogenic testbed at NASA Ames Research Center to assess the likely measurement precision of a future space telescope that would make these observations. The testbed utilizes the same blackbody source to simulate a star and has a densified pupil spectrograph with the JWST prototype mid-infrared Si:As IBC 1024 x 1024 pixel detector. This performance acquired in this study will serve as a performance baseline when we measure the entire system performance including measurement precision at a later date. 

\acknowledgements
We thank George Rieke for providing detector operation insights which improved this work. T. Matsuo is sincerely grateful to the support of Osaka University, NASA Ames Research Center, and NASA Headquaters. This work was supported by JSPS KAKENHI Grant Numbers, 16H02164, 17KK0090, 19H00700 and the Mitsubishi Foundation. T. Greene acknowledges support by NASA Ames Research Center internal research and development (IRAD) funding. Finally, we express our sincere gratitude to the referee for reading our draft in detail and carefully checking the values.


\begin{thebibliography}{}
\bibitem[Agol et al.(2010)]{2010ApJ...721.1861A} Agol, E., Cowan, N. B., Knutson, H. A., et al. \ 2010, \apj, 721, 1861
\bibitem[Barstow et al.(2016)]{2016MNRAS.458.2657B} Barstow, J. K., Aigrain, S., Irwin, P. G. J., Kendrew, S, Fletcher, L. N. \ 2016, \mnras, 458, 2657
\bibitem[Barstow \& Irwin (2016)]{2016MNRAS.461L..92B} Barstow, J. K. \& Irwin, P. G. J. \ 2016, \mnras, 461L, 92
\bibitem[Beichman et al. (2014)]{2014PASP..126.1134B} Beichman, C., Benneke, B., Knutson, H., et al. \ 2014, \pasp, 126, 1134
\bibitem[Ennico et al.(2003)]{2003SPIE.4850..890E} Ennico, K. A., McKelvey, M. E., McCreight, C. R., et al.\ 2003, Proceedings of the SPIE, 4850, 890-901
\bibitem[Goda \& Matsuo (2018)]{2018AJ....156..288G} Goda, S. \& Matsuo, T. \ 2018, \aj, 156, 288
\bibitem[Greene et al. (2016)]{2016ApJ...817...17G} Greene, T. P., Line, M. R., Montero, C., et al. \ 2016, \apj, 817, 17 
\bibitem[Hora et al.(2004)]{2004SPIE.5487...77H} Hora, J. L., Fazio, G. G., Allen, L. E., et al. \ 2004, Proceedings of the SPIE, 5487, 77-92
\bibitem[Ingalls et al.(2016)]{2016AJ....152...44I} Ingalls, J. G., Krick, J. E., Carey, S. J., Stauffer, J. R., et al. \ 2016, \aj, 152, 44
\bibitem[Itoh et al.(2017)]{2017AJ....154...97I} Itoh, S., Matsuo, T., Goda, S., Shibai, H. Sumi, T. \ 2017, \aj, 154, 97
\bibitem[Kilpatrick et al.(2019)]{2019arXiv190402294K} Kilpatrick, B. M., Kataria, T., Lewis, N. K., et al.\ 2019, arXiv: 1904.02994
\bibitem[Knutson et al.(2009)]{2009ApJ...703..769K} Knutson, H. A., Charbonneau, D., Cowan, N. B., et al.\ 2009, \apj, 703, 769
\bibitem[Knutson et al.(2012)]{2012ApJ...754...22K} Knutson, H. A., Lewis, N., Fortney, J. J., et al.\ 2012, \apj, 754, 22
\bibitem[Kreidberg et al.(2014)]{2014Natur.505...69K} Kreidberg, L., Bean, J. L., D{\'e}sert, J.-M., et al.\ 2014, Nature, 505, 69-72
\bibitem[Mandel \& Agol(2002)]{2002ApJ...580L.171M} Mandel, K., \& Agol, E. \ 2002, \apj, 580L, 171-175
\bibitem[Matsuo et al.(2016)]{2016ApJ...823..139M} Matsuo, T., Itoh, S., Shibai, H., Sumi, T., Yamamuro, T.\ 2016, \apj, 823, 139
\bibitem[Matsuo et al.(2018)]{2018SPIE10698E..44M} Matsuo, T., Greene, T., Roellig, T. L., et al. \ 2018, Proceedings of the SPIE, 10698, 1069844
\bibitem[Morello et al.(2016)]{2016ApJ...820...86M} Morello, G., Waldmann, I. P., \& Tinetti, G. \ 2016, \apj, 820, 86
\bibitem[Ressler et al.(2015)]{2015PASP..127..675R} Ressler, M. E., Sukhatme, K. G., Franklin, B. R., et al. \ 2015, \pasp, 127, 675
\bibitem[Rieke et al.(2015)]{2015PASP..127..665R} Rieke, G. H., Ressler, M. E., Morrison, J. E., et al. \ 2015, \pasp, 127, 665
\bibitem[Schwieterman et al.(2016)]{2016ApJ...819L..13S} Schwieterman, E. W., Meadows, V. S., Domagal-Goldman, S. D., et al. \ 2016, \apj, 819L, 13


\end{thebibliography}
\end{document}